%% file: bare_jrnl.tex
\documentclass[journal]{IEEEtran}

\ifCLASSINFOpdf
\else
   \usepackage[dvips]{graphicx}
\fi
\usepackage{url}

\usepackage{graphicx}
\usepackage{booktabs}
\usepackage[table,xcdraw]{xcolor}
\usepackage{colortbl}
\definecolor{Gray}{gray}{0.9}
\usepackage{cite}           
\usepackage{amsmath}
\usepackage{amssymb}
\usepackage{pifont}
\newcommand{\cmark}{\ding{51}}%
\newcommand{\xmark}{\ding{55}}%
\usepackage{multirow}
\usepackage[caption=false]{subfig}
\usepackage{hyperref}
\definecolor{blue}{RGB}{0,0,255}

\begin{document}

\title{mWhisper-Flamingo for Multilingual Audio-Visual Noise-Robust Speech Recognition}

\author{Andrew Rouditchenko,
        Samuel Thomas,~\IEEEmembership{Senior Member, IEEE},
        Hilde Kuehne,~\IEEEmembership{Member, IEEE},
        Rogerio Feris,~\IEEEmembership{Senior Member, IEEE},
        and James Glass,~\IEEEmembership{Fellow, IEEE}
\thanks{Code at \href{https://github.com/roudimit/whisper-flamingo}
{\texttt{https://github.com/roudimit/whisper-flamingo}}.
Andrew Rouditchenko and James Glass are with MIT, USA (e-mail: \{roudi, glass\}@mit.edu).
Samuel Thomas and Rogerio
Feris are with MIT-IBM Watson AI Lab, USA.
Hilde Kuehne is with University of Tuebingen, DE.
We thank Videet Mehta for help with the demo and Tatiana Likhomanenko and Saurabhchand Bhati for helpful discussions. This research was supported by MIT-IBM Watson AI Lab and an NDSEG Fellowship to A.R.}}

\markboth{IEEE Signal Processing Letters}
{Shell \MakeLowercase{\textit{et al.}}: Bare Demo of IEEEtran.cls for IEEE Journals}
\maketitle

\begin{abstract}
Audio-Visual Speech Recognition (AVSR) combines lip-based video with audio and can improve performance in noise, but most methods are trained only on English data.
One limitation is the lack of large-scale multilingual video data, which makes it hard to train models from scratch.
In this work, we propose mWhisper-Flamingo for multilingual AVSR which combines the strengths of a pre-trained audio model (Whisper) and video model (AV-HuBERT).
To enable better multi-modal integration and improve the noisy multilingual performance, we introduce decoder modality dropout where the model is trained both on paired audio-visual inputs and separate audio/visual inputs.
mWhisper-Flamingo achieves state-of-the-art WER on MuAViC, an AVSR dataset of 9 languages.
Audio-visual mWhisper-Flamingo consistently outperforms audio-only Whisper on all languages in noisy conditions.
\end{abstract}


\IEEEpeerreviewmaketitle

\section{Introduction}
Automatic Speech Recognition (ASR) has seen great progress thanks to large-scale training~\cite{radford2023robust,puvvada24_interspeech,rouditchenko23_interspeech,shi23g_interspeech}, but models still struggle with background noise~\cite{gong23d_interspeech,hu2024large}.
To improve performance, Audio-Visual Speech Recognition (AVSR) models combine lip-based video with audio inputs~\cite{afouras2018adeep,petridis2018audio,petridis2018end,xu2020discriminative,ma2021end,serdyuk22_interspeech,shi22_interspeech,ma2023auto,burchi2023audio,cappellazzo2024large}.
In particular, Whisper-Flamingo~\cite{rouditchenko24_interspeech} proposed an audio-visual adaptation of Whisper~\cite{radford2023robust}, a pre-trained ASR model, and showed significant improvements in noise robustness compared to the original audio-only model.
Whisper-Flamingo was motivated by the limitation of previous AVSR systems which often lack large-scale transcribed videos and face difficulty training models from scratch on only a few hundred hours of data.
To overcome this, Whisper-Flamingo combines the strength of Whisper's audio encoder and text decoder trained on 680k hours with AV-HuBERT~\cite{shi2022learning}, a pre-trained lip-reading visual encoder.
The model integrates lip-based visual features from AV-HuBERT into Whisper's decoder and achieves State-of-the-Art (SOTA) performance on English AVSR.

In this work, we propose mWhisper-Flamingo: a novel multilingual extension of Whisper-Flamingo which achieves SOTA performance on multilingual AVSR.
mWhisper-Flamingo combines Whisper's strong multilingual audio encoder and text decoder with a new AV-HuBERT visual encoder pre-trained on multilingual videos~\cite{kim_2024}.
Unlike the previous Whisper-Flamingo model which could only take English videos as input, the new mWhisper-Flamingo model can handle videos in 9 different languages (including English).
However, we show that Whisper-Flamingo's default training process applied to multilingual videos yields poor noisy multilingual AVSR performance, despite achieving good English performance.
To address this, we propose a novel decoder modality dropout technique by training the model both on paired audio-visual inputs and separate audio/video inputs.
We show this to be key for good noisy multilingual AVSR performance with a thorough analysis and ablation study.

We test our method across 9 languages in clean and noisy conditions on the MuAViC dataset~\cite{anwar23_interspeech}.
In clean audio conditions, mWhisper-Flamingo outperforms previous audio-visual methods and achieves SOTA across multilingual languages.
In noisy conditions, mWhisper-Flamingo consistently outperforms the audio-only Whisper model on 6 different noise types and 5 levels of noise.
We release our code and models.

\begin{figure}[t]
    \centering
    \includegraphics[width=\linewidth]{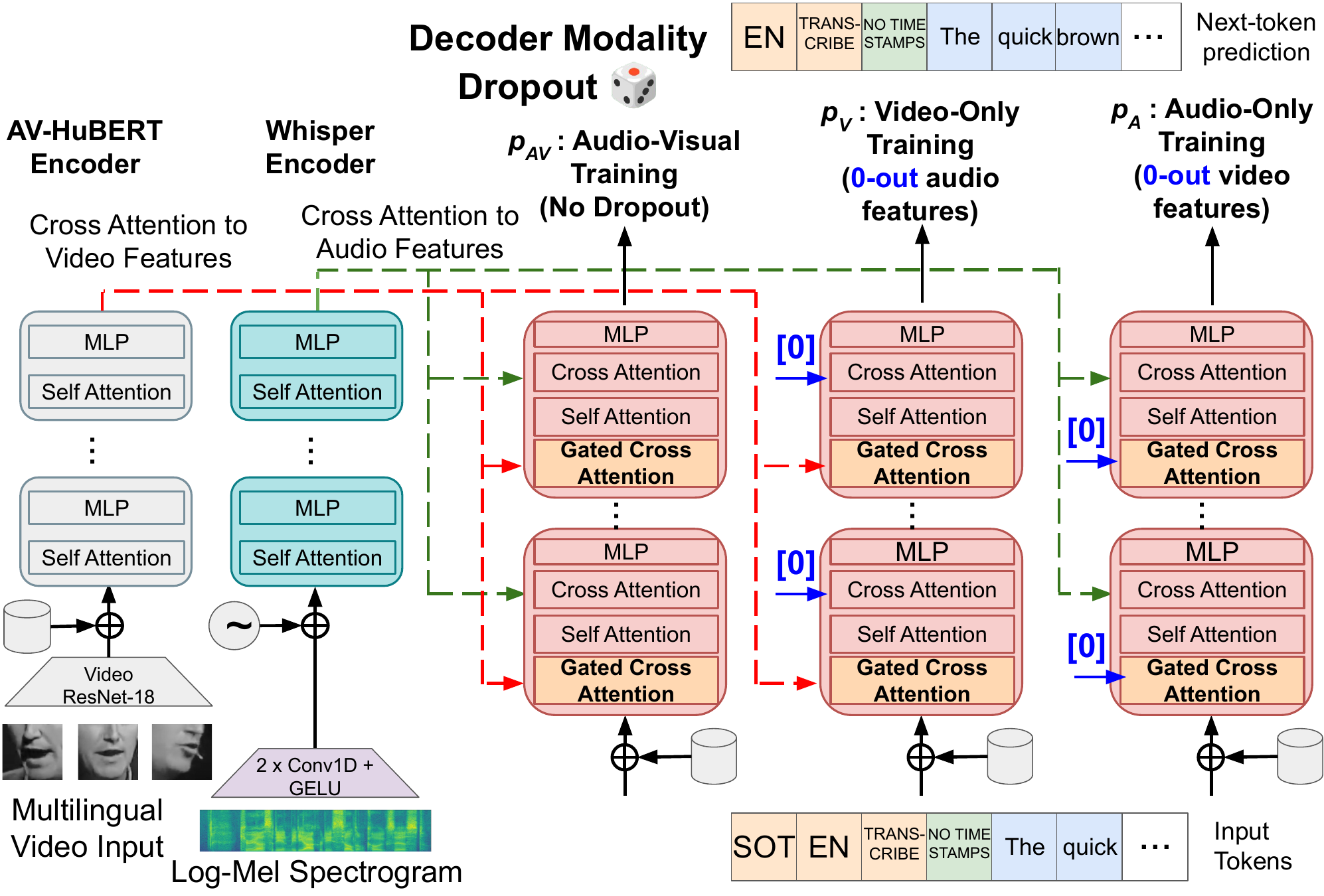}
    \caption{In mWhisper-Flamingo, the AV-HuBERT and Whisper encoders extract visual and audio features from multilingual videos. 
    Separate cross attention layers in Whisper's decoder attend to the visual and audio features.
    Decoder modality dropout randomly replaces the audio or video features by 0, forcing the decoder to train on video-only and audio-only inputs.}
    \label{fig:overview}
    \vspace{-0.4cm}
\end{figure}

\section{Method}
Our method builds upon Whisper-Flamingo~\cite{rouditchenko24_interspeech}, an audio-visual extension of Whisper~\cite{radford2023robust}.
Whisper-Flamingo adds new gated cross-attention layers (originally proposed for the Flamingo vision-language model~\cite{alayrac2022flamingo}) into each of Whisper’s decoder blocks which attend to the visual features from the AV-HuBERT visual encoder~\cite{shi2022learning}.
The layers are initialized as identity functions and initially ignore the visual input, but the weights are adjusted to integrate the visual features for AVSR during training on audio-visual inputs.

Whisper-Flamingo uses two-stages of training.
First, all of Whisper's parameters are fine-tuned on noisy audio inputs, enhancing its domain-specific performance and noise robustness. 
Second, the gated cross-attention layers are initialized and trained on audio-visual data. 
During this stage, all of Whisper's and AV-HuBERT's parameters are frozen to preserve the pre-trained knowledge and to facilitate multi-modal integration.

mWhisper-Flamingo inherits Whisper-Flamingo's architecture, except the English AV-HuBERT~\cite{shi2022learning} is replaced with a version pre-trained on multilingual videos~\cite{kim_2024}.
We propose to train mWhisper-Flamingo on multilingual videos jointly with English, similar to Whisper's training process.
However, using Whisper-Flamingo's default training process on multilingual videos yielded only minor improvements in noisy multilingual AVSR performance, despite achieving significant improvements for English (see Section~\ref{sec:analysis}). 
We suspect this is due to English having on average 13.6x more data than other languages in our dataset.
To address this, we introduce decoder modality dropout by training the model both on paired audio-visual inputs and separate audio/video inputs.

Dropout, originally proposed to prevent overfitting in neural networks~\cite{hinton2012improving}, was extended to multi-modal learning as modality dropout~\cite{neverova2015moddrop}. 
Modality dropout randomly drops input modalities during training to better capture cross-modality correlations while preserving unique contributions of specific modalities.
Modality dropout has been used in AVSR~\cite{makino2019recurrent,shi2022learning,hsu2022u,lian2023av,rouditchenko2023av} to prevent models from over-relying on the audio modality, which is typically easier to transcribe than the lip-based visual modality.
However, current methods mainly use early-fusion where audio and visual features are combined and used as input into a single Transformer~\cite{vaswani2017attention} encoder.
During modality dropout, the Transformer encoder must handle the incomplete input stream and output an embedding sequence which is used as input to the decoder's cross-attention layers.
In contrast, Whisper-Flamingo uses late-fusion where separate Transformer encoders process audio and visual features independently.
Each encoder outputs an embedding sequence used as input to separate audio and video cross attention layers in the decoder, where the modality fusion occurs.
During modality dropout, one of the audio and video embedding sequences is replaced by a 0-vector, forcing the decoder to handle the incomplete input stream in each of the cross-attention layers corresponding to the missing modality.
This enables the encoders to remain specialized on their specific modalities while the decoder learns better multi-modal integration.
This process is shown in Figure~\ref{fig:overview}.
Note that cross-attention with a 0-vector results in a 0-vector, but the output from the layer incorporates the bias in the linear transformations.

Our method is conceptually similar to LayerDrop~\cite{Fan2020Reducing}, which randomly drops Transformer layers for model pruning on text-based tasks. 
However, our focus lies in enhancing the decoder's ability to learn better multi-modal integration and to handle inputs where one modality may be unreliable.

We define the probabilities of using audio-visual (\( p_{AV} \)), audio-only (\( p_{A} \)), and visual-only (\( p_{V} \)) inputs during training as follows. 
If audio-visual inputs are selected, both modalities are used.  
If audio-only inputs are selected, visual features are zeroed out at the decoder.  
If video-only inputs are selected, audio features are zeroed out at the decoder.  
In Section~\ref{sec:analysis}, we show results using different probabilities.
The setting \( p_{AV} = 0.5,\ p_{A} = 0,\ p_{V} = 0.5 \) performs the best on noisy multilingual AVSR, so we use it for training our main models.

\section{Experiments}

\subsection{Experimental Setup}

We use the MuAViC~\cite{anwar23_interspeech} dataset of 1,141h of videos in 9 languages.
The dataset is based on the LRS3 English video dataset~\cite{afouras2018lrs3} and mTedX dataset~\cite{salesky21_interspeech}.
The hours of video per language are: 
English (En): 433, Arabic (Ar): 16, German (De): 10, Greek (El): 25, Spanish (Es): 178, French (Fr): 176, Italian (It): 101, Portuguese (Pt): 153 and Russian (Ru): 49.

We use Whisper~\cite{radford2023robust} small, medium, and large-v2 with 244M, 769M, and 1.55B parameters.
We fine-tuned Whisper small and medium on 4 A6000 GPUs with 48GB memory, but could not fine-tune Whisper large due to the GPU memory limits.
We use AV-HuBERT pre-trained on unlabeled multilingual videos~\cite{kim_2024} with 325M parameters.
We selected this model instead of other English-only lip-reading models~\cite{haliassos2023jointly,haliassos2024braven,haliassos2024unified} due to its superiority on multilingual videos~\cite{ma2022visual,zinonos2023learning,kim2023lip,yeo2024visual}.
Different to Whisper-Flamingo, we also fine-tune its parameters.
The gated cross attention layers add 82M and 296M for the small and medium models, bringing the total to 651M and 1.39B parameters for mWhisper-Flamingo small / medium. 
The other dataloading details and hyperparameters closely follow Whisper-Flamingo~\cite{rouditchenko2023av}.
Spectrogram frames are used as input to Whisper at 100 Hz while grayscale videos are used as input to AV-HuBERT at 25 fps.
The videos are cropped on the lips using Dlib~\cite{king2009dlib} and are aligned to a reference mean face~\cite{martinez2020lipreading}.
Models were trained using PyTorch~\cite{paszke2019pytorch} and PyTorch Lightning~\cite{Falcon_PyTorch_Lightning_2019} with the AdamW optimizer~\cite{loshchilov2018decoupled}.

During training, we randomly add noise to the audio with a Signal-to-Noise Ratio (SNR) of 0 dB. 
Based on prior work~\cite{shi22_interspeech,anwar23_interspeech}, ``natural'', ``music'' and ``babble'' noise are sampled from the MUSAN dataset~\cite{snyder2015musan}, and overlapping ``speech'' is sampled from LRS3~\cite{afouras2018lrs3}.
We monitor the token prediction accuracy on the noisy validation set every 1k steps to select the best checkpoints.
We normalize the text by lower-casing and removing punctuation except single apostrophes.
Our goal is to improve the multilingual Word Error Rate (WER), so we compute average WER on all languages except English. 
Given the data imbalance, we also separately compute the average on the ``higher resource'' languages with over 100h of data (Es, Fr, It, Pt), and the other ``lower resource'' languages (Ar, De, El, Ru) with as low as 10h of data.

\input{tables/main}
\input{tables/noisy}

\subsection{Clean Results}
In Table~\ref{tab:main}, we show the results on MuAViC using the original, clean audio.
For previous SOTA audio-visual methods, initial work fine-tuned pre-trained English AV-HuBERT~\cite{shi2022learning} on multilingual videos in MuAViC~\cite{anwar23_interspeech,hong-etal-2023-intuitive}.
Some methods trained individual models for specific non-En languages~\cite{li2023parameter,gimeno2024tailored,li24i_interspeech}, however, they are outperformed by multilingual models.
The current SOTA models are XLAVS-R~\cite{han2024xlavs} and Fast Conformer~\cite{burchi2024multilingual}.
The former adapted XLS-R~\cite{babu2021xls}, a multilingual self-supervised audio model, and the latter trains a model from scratch. 
Moreover, Fast Conformer achieved even better results on the higher-resource languages by using extra multilingual training videos from other datasets.
While these methods worked well on the higher-resource languages, their performance on lower-resource languages is less satisfactory.

Next, we establish baselines using Whisper.
Compared to XLAVS-R 2B, Whisper small zero-shot (without fine-tuning) achieves better performance on the lower resource languages (37.5\% vs 41.9\%), while Whisper medium zero-shot achieves better overall average non-En WER (22.9\% vs 26.4\%).
This shows Whisper's strong multilingual capabilities and motivates its selection as the foundation of our proposed models.

Fine-tuning Whisper on MuAViC leads to further gains.
Fine-tuned Whisper medium achieves a new \textbf{SOTA non-En average WER of 20.1\%,} surpassing all previous audio-visual methods fine-tuning solely on MuAViC. 
Notably, the English performance is also improved from 2.3\% (zero-shot) to 0.74\%, approaching the current SOTA of 0.68\% reported in Whisper-Flamingo~\cite{rouditchenko24_interspeech}. 
This shows that fine-tuning improves multilingual performance without compromising English accuracy.  

mWhisper-Flamingo performs similarly to audio-only fine-tuned Whisper (20.4\% vs. 20.1\% WER for the medium models). 
This small difference suggests that video provides limited benefit for clean audio. 
However, mWhisper-Flamingo still significantly outperforms all previous AVSR models fine-tuned solely on the 1,141h MuAViC videos, achieving a new SOTA on all languages with this setup. 
It even surpasses Fast Conformer trained with 4,957h of videos on all languages except Fr and Pt. 
This shows the strength of using Whisper as initialization for AVSR. 
Finally, mWhisper-Flamingo medium consistently outperforms mWhisper-Flamingo small, showing the benefit of increased model size.

\subsection{Noisy Results}
Table~\ref{tab:noisy} shows the performance on MuAViC with babble noise at 0-SNR. 
The babble noise is from Whisper-Flamingo~\cite{rouditchenko24_interspeech} and was constructed from 30 LRS3 speakers.
We compare Whisper zero-shot with Whisper fine-tuned and mWhisper-Flamingo.
The noisy results show a significant performance degradation compared to the clean results (Table~\ref{tab:main}).
For example, the average non-En WER for Whisper small zero-shot increases from 27.0\% in the clean setting to 71.0\% under 0-SNR babble noise. 
However, fine-tuning significantly improves WER. 
For the small model, fine-tuning improves the average non-En WER from 71.0\% (zero-shot) to 55.3\%. 
A similar trend is observed for the medium model, with WER decreasing from 60.7\% to 48.0\%. 

The integration of visual features in mWhisper-Flamingo provides further gains. 
mWhisper-Flamingo small achieves an average non-En WER of 50.4\%, a \textbf{10.4\% relative improvement} over fine-tuned audio-only Whisper small (55.3\%).
Similarly, mWhisper-Flamingo medium achieves an average non-En WER of 43.7\%, a \textbf{10.6\% relative improvement} compared to fine-tuned audio-only Whisper medium (48.0\%). 
The relative improvements are better for the higher-resource languages (\textbf{16.0\% and 16.4\%} for small and medium models), indicating that more video training data is helpful.
The relative improvements for English are much better at 48.4\% and 39.8\% for the small and medium models, which we attribute to having more English training data.

mWhisper-Flamingo medium achieves the best multilingual WER (43.7\%), even outperforming Whisper large zero-shot (56.0\%) despite having fewer parameters (1.39B vs 1.5B).
This shows that the performance gains are not only due to more parameters but also due to the visual modality. 
As additional evidence, mWhisper-Flamingo small (651M parameters) outperforms audio-only fine-tuned Whisper medium (769M parameters) on the higher-resource languages (37.4\% vs. 38.1\%). 
This highlights that smaller audio-visual models can achieve better performance than larger audio-only models.

Finally, we tested the models on 6 different noise types, 5 SNR levels $\{-10,-5,0,5,10 \}$, and 4 languages (Es, Fr, It, Pt). 
The noise setups follow Whisper-Flamingo~\cite{rouditchenko24_interspeech} and AV-HuBERT~\cite{shi22_interspeech}.
Figure~\ref{fig:noise} shows the WER for Es, Fr, It, Pt averaged over all SNR values for each noise type. 
There is a clear trend of mWhisper-Flamingo outperforming the audio-only models, especially for the more challenging noises (3 types of babble and side speech).
Overall, these results confirm the advantage of mWhisper-Flamingo on diverse noise types.

\input{figures/noise_fig}

\subsection{Ablation Study and Analysis}
\label{sec:analysis}
In Table~\ref{tab:ablation}, we present an ablation study and model analysis.
Due to computational constraints, we trained and evaluated the small model on 5 of the 9 languages.
We compare the average non-En WER between models.
We show the noisy results since the performance differences between models in the clean results were minor.
The baseline audio-only fine-tuned Whisper small achieves 44.4\%.

\input{tables/ablation}
First, we show that the combination of decoder modality dropout and a fine-tunable visual encoder leads to the best performance (36.6\%).
The performance is significantly worse if decoder modality dropout is disabled using either a fine-tunable visual encoder (44.6\%) or a frozen visual encoder encoder (42.6\%).
It shows that our proposed decoder modality dropout is \textit{crucial} for obtaining the best multilingual performance.
Note that the latter setup represents the default training configuration from Whisper-Flamingo~\cite{rouditchenko24_interspeech}.
Considering that the latter setup improved the English noisy WER from 16.2\% to 11.5\%, it is reasonable that modality dropout was not necessary in the original, English-only Whisper-Flamingo.
Finally, using decoder modality dropout with a fine-tunable visual encoder is better than using it with a frozen visual encoder (36.6\% vs 40.6\%).

mWhisper-Flamingo uses decoder modality dropout probabilities of \( p_{AV} = 0.5,\ p_{A} = 0,\ p_{V} = 0.5 \), meaning that the model trains with audio-visual inputs 50\% of the time and visual-only inputs 50\% of the time.
Without dropout, the modal trains only on audio-visual inputs (\( p_{AV} = 1,\ p_{A} = 0,\ p_{V} = 0 \)), and the performance is much worse (44.6\% vs 36.6\%).
Using dropout with audio-only inputs 50\% of the time instead of video-only inputs (\( p_{AV} = 0.5,\ p_{A} = 0.5,\ p_{V} = 0 \)) does not improve performance (44.6\%). 
It shows that it is necessary for the model to train with video-only inputs.
Trying different probabilities between AV, A, and V such as \( p_{AV} = 0.5,\ p_{A} = 0.25,\ p_{V} = 0.25 \) and \( p_{AV} = 0.25,\ p_{A} = 0.25,\ p_{V} = 0.5 \), the performance is improved to 37.7\%, however the best performance is achieved without audio-only inputs (\( p_{AV} = 0.5,\ p_{A} = 0,\ p_{V} = 0.5 \)).
Our explanation is that Whisper was fine-tuned in the first stage of training, so the model can already handle audio-only inputs well.
Training on video-only inputs allows the model to better integrate the visual modality.
We also tried other proportions of audio-visual and video-only training such as \( p_{AV} = 0.75,\ p_{A} = 0,\ p_{V} = 0.25 \) and \( p_{AV} = 0.25,\ p_{A} = 0,\ p_{V} = 0.75 \) which performed slightly worse.

Finally, we compare multilingual AV-HuBERT~\cite{kim_2024} with AV-HuBERT pre-trained only on English videos~\cite{shi2022learning}.
The model using the multilingual AV-HuBERT encoder achieves better multilingual performance (36.6\% vs 37.4\%) but worse English performance (8.0\% vs 7.5\%), which is reasonable.

\section{Conclusion}
We introduce mWhisper-Flamingo, a novel multilingual model for AVSR.
mWhisper-Flamingo outperforms all audio-visual methods trained on MuAViC, and even surpasses a model trained with substantially more videos on most languages. 
mWhisper-Flamingo outperforms audio-only Whisper in diverse noise settings. 
Our proposed decoder modality dropout significantly improved the noisy multilingual WER.

\clearpage

\bibliographystyle{IEEEtran}
\bibliography{ref}

\setcounter{table}{0}
\renewcommand{\thetable}{A\arabic{table}}
\section{Appendix}
\subsection{Full Noisy Results}
\input{tables/noise_full_small}

\input{tables/noise_full_medium}

Table~\ref{tab:noise-full-small} and Table~\ref{tab:noise-full-medium} show the full decoding results for different noise types and SNRs.
Table~\ref{tab:noise-full-small} compares the small models and Table~\ref{tab:noise-full-medium} compares the medium models.
Note that the results in Table~\ref{tab:noise-full-small} were used to create Figure~\ref{fig:noise} (excluding the English results).

\end{document}

%% file: tables/main.tex
\begin{table*}[t]
    \centering
    \setlength{\tabcolsep}{8pt}
    \caption{Multilingual WER on MuAViC (clean).
    $\downarrow$ is better.
    Hours of video used to fine-tune each model are shown. 
    Mod. = Modality.
    Avg. non-En: average WER without English.
    H.R. = High Resource (Es, Fr, It, Pt). L.R. = Low Resource (Ar, De, El, Ru).
    }
    \label{tab:main}
    \vspace{-3mm}
\resizebox{\linewidth}{!}{%
\begin{tabular}{lcccrrrrrrrrrrrr}
\toprule
Model & \begin{tabular}[c]{@{}l@{}}Total \\ Params\end{tabular} & \begin{tabular}[c]{@{}l@{}}FT Vid. \\ Hours\end{tabular} & Mod. & \multicolumn{1}{l}{En} & \multicolumn{1}{l}{Ar} & \multicolumn{1}{l}{De} & \multicolumn{1}{l}{El} & \multicolumn{1}{l}{Es} & \multicolumn{1}{l}{Fr} & \multicolumn{1}{l}{It} & \multicolumn{1}{l}{Pt} & \multicolumn{1}{l}{Ru} & \multicolumn{1}{l}{\begin{tabular}[c]{@{}l@{}}Avg \\ non-En\end{tabular}} & \multicolumn{1}{l}{\begin{tabular}[c]{@{}l@{}}Avg \\ H.R.\end{tabular}} & \multicolumn{1}{l}{\begin{tabular}[c]{@{}l@{}}Avg\\ L.R. \end{tabular}} \\


\hline
\rowcolor{Gray}\multicolumn{16}{c}{\textit{Multilingual Audio-Visual Baselines}} \\
AV-HuBERT~\cite{shi2022learning,han2024xlavs} & 477M & 1,141& AV & \bf{1.7} & 89.4 & 52.0 & 46.2 & 17.4 & 20.3 & 20.8 & 22.1 & 44.7 & 39.1 & 20.2 & \bf{58.1} \\
Intuitive Multilingual~\cite{hong-etal-2023-intuitive} & 477M & 1,141& AV & 2.5 & 90.9 & 48.1 & 42.4 & 18.0 & 21.5 & 19.7 & 20.8 & 54.1 & 39.5 & 20.0 & 58.9 \\
XLAVS-R 300M~\cite{han2024xlavs} & 452M & 1,141& AV & 2.4 & 81.7 & 44.7 & 24.3 & 10.9 & 14.4 & 12.8 & 13.2 & 32.7 & 29.3 & 12.8 & 45.9 \\
XLAVS-R 2B~\cite{han2024xlavs} & 2.15B & 1,141& AV & \bf{1.7} & \bf{79.3} & \bf{44.4} & \bf{19.0} & 9.1 & 12.3 & \bf{10.6} & \bf{11.2} & \bf{25.0} & \bf{26.4} & \bf{10.8} & \bf{41.9} \\
Fast Conformer~\cite{burchi2024multilingual} & 197M & 1,141& AV & - & - & 46.8 & - & \bf{9.0} & \bf{11.4} & 11.6 & 11.8 & - & - & 11.0 & - \\

\hline
\rowcolor{Gray}\multicolumn{16}{c}{\textit{Multilingual Audio-Visual Baselines (Using Extra Multilingual Training Videos)}} \\
Fast Conformer~\cite{burchi2024multilingual} & 197M & 4,957 & AV & - & - & \bf{23.6} & - & \bf{8.2} & \bf{10.3} & \bf{10.4} & \bf{9.9} & - & - & \bf{9.7} & - \\

\hline
\rowcolor{Gray}\multicolumn{16}{c}{\textit{Our Audio-Only Whisper Zero-Shot Baselines}} \\
Whisper Small Zero-Shot & 244M & - & A & 2.6 & 78.9 & 23.9 & 23.5 & 11.4 & 17.7 & 20.0 & 17.1 & 23.7 & 27.0 & 16.6 & 37.5 \\
Whisper Medium Zero-Shot & 769M & - & A & 2.3 & 75.6 & 21.3 & 15.7 & 9.5 & 15.6 & 11.6 & 13.1 & 20.7 & 22.9 & 12.5 & 33.3 \\
Whisper Large Zero-Shot & 1.5B & - & A & \bf{2.1} & \bf{73.2} & \bf{20.1} & \bf{12.7} & \bf{8.9} & \bf{14.5} & \bf{10.2} & \bf{11.7} & \bf{19.0} & \bf{21.3} & \bf{11.3} & \bf{31.3} \\

\hline
\rowcolor{Gray}\multicolumn{16}{c}{\textit{Our Models}} \\
Whisper Small Fine-Tuned & 244M & - & A & 1.1 & 73.3 & 26.4 & 18.4 & 9.5 & 13.8 & 12.8 & 12.8 & 21.2 & 23.5 & 12.2 & 34.8 \\
mWhisper-Flamingo Small & 651M & 1,141& AV & 1.2 & 74.9 & 26.6 & 18.9 & 9.6 & 13.8 & 12.7 & 12.9 & 21.2 & 23.8 & 12.3 & 35.4 \\
Whisper Medium Fine-Tuned & 769M & - & A & \bf{0.74} & \bf{68.9} & \bf{21.5} & \bf{13.6} & \bf{7.9} & \bf{11.3} & \bf{10.1} & \bf{10.7} & \bf{16.7} & \bf{20.1} & \bf{10.0} & \bf{30.2} \\
mWhisper-Flamingo Medium & 1.39B & 1,141& AV & 0.88 & 70.2 & 22.0 & 14.0 & 8.1 & \bf{11.3} & 10.2 & 10.8 & \bf{16.7} & 20.4 & 10.1 & 30.7\\
\bottomrule
\end{tabular}%
}
\vspace{-0.4cm}
\end{table*}

%% file: tables/noisy.tex
\begin{table*}[t]
    \centering
    \setlength{\tabcolsep}{8pt}
    \caption{
    Multilingual WER on MuAViC (Babble noise added at 0-SNR).
    Hours of video used to fine-tune each model are shown. 
    Mod. = Modality.
    Avg. non-En: average WER without English.
    H.R. = High Resource (Es, Fr, It, Pt). L.R. = Low Resource (Ar, De, El, Ru).
    }
    \label{tab:noisy}
    \vspace{-3mm}
\resizebox{\linewidth}{!}{%
\begin{tabular}{lcccrrrrrrrrrrrr}
\toprule
Model & \begin{tabular}[c]{@{}l@{}}Total \\ Params\end{tabular} & \begin{tabular}[c]{@{}l@{}}FT Vid. \\ Hours\end{tabular} & Mod. & \multicolumn{1}{l}{En} & \multicolumn{1}{l}{Ar} & \multicolumn{1}{l}{De} & \multicolumn{1}{l}{El} & \multicolumn{1}{l}{Es} & \multicolumn{1}{l}{Fr} & \multicolumn{1}{l}{It} & \multicolumn{1}{l}{Pt} & \multicolumn{1}{l}{Ru} & \multicolumn{1}{l}{\begin{tabular}[c]{@{}l@{}}Avg \\ non-En\end{tabular}} & \multicolumn{1}{l}{\begin{tabular}[c]{@{}l@{}}Avg \\ H.R.\end{tabular}} & \multicolumn{1}{l}{\begin{tabular}[c]{@{}l@{}}Avg\\ L.R. \end{tabular}} \\
\hline
\rowcolor{Gray}\multicolumn{16}{c}{\it{Small Models}} \\
Whisper Small Zero-Shot & 244M & - & A & 27.9 & 102 & 61.2 & 76.4 & 63.3 & 57.8 & 76.2 & 73.4 & 57.6 & 71.0 & 67.7 & 74.3 \\
Whisper Small Fine-Tuned & 244M & - & A & 16.1 & \bf{101} & 59.9 & 56.8 & 41.7 & 35.9 & 50.6 & 50.2 & 46.7 & 55.3 & 44.6 & 66.0 \\
mWhisper-Flamingo Small & 651M & \multicolumn{1}{r}{1,141} & AV & \bf{8.3} & \bf{101} & \bf{56.7} & \bf{51.1} & \bf{33.6} & \bf{31.9} & \bf{41.2} & \bf{42.8} & \bf{44.6} & \bf{50.4} & \bf{37.4} & \bf{63.5} \\
Relative Improvement &  & - &  & \it{48.4} & \it{-0.8} & \it{5.3} & \it{10.0} & \it{19.4} & \it{11.1} & \it{18.6} & \it{14.7} & \it{4.5} & \it{10.4} & \it{16.0} & \it{4.8} \\
\hline
\rowcolor{Gray}\multicolumn{16}{c}{\it{Medium Models}} \\
Whisper Medium Zero-Shot & 769M & - & A & 22.5 & 105 & 53.1 & 61.3 & 49.1 & 47.7 & 60.3 & 60.9 & 47.5 & 60.7 & 54.5 & 66.8 \\
Whisper Medium Fine-Tuned & 769M & - & A & 12.3 & 96.4 & 51.8 & 45.7 & 36.1 & 30.4 & 43.5 & 42.2 & 37.9 & 48.0 & 38.1 & 58.0 \\
mWhisper-Flamingo Medium & 1.39B & \multicolumn{1}{r}{1,141} & AV & \bf{7.4} & \bf{95.3} & \bf{49.4} & \bf{41.8} & \bf{28} & \bf{27.5} & \bf{35.2} & \bf{36} & \bf{36.1} & \bf{43.7} & \bf{31.7} & \bf{55.7} \\
Relative Improvement &  & - &  & \it{39.8} & \it{1.1} & \it{4.6} & \it{8.5} & \it{22.4} & \it{9.5} & \it{19.1} & \it{14.7} & \it{4.7} & \it{10.6} & \it{16.4} & \it{4.8} \\
\hline
\rowcolor{Gray}\multicolumn{16}{c}{\it{Large Models}} \\
Whisper Large Zero-Shot & 1.5B & - & A & \bf{21.3} & \bf{96.5} & \bf{51.5} & \bf{54.7} & \bf{46.9} & \bf{41.7} & \bf{56.6} & \bf{58.7} & \bf{41.3} & \bf{56.0} & \bf{51.0} & \bf{61.0}\\
\bottomrule
\end{tabular}%
}
\vspace{-0.4cm}
\end{table*}

%% file: figures/noise_fig.tex
\begin{figure}[t]
    \centering
    \includegraphics[width=\linewidth]{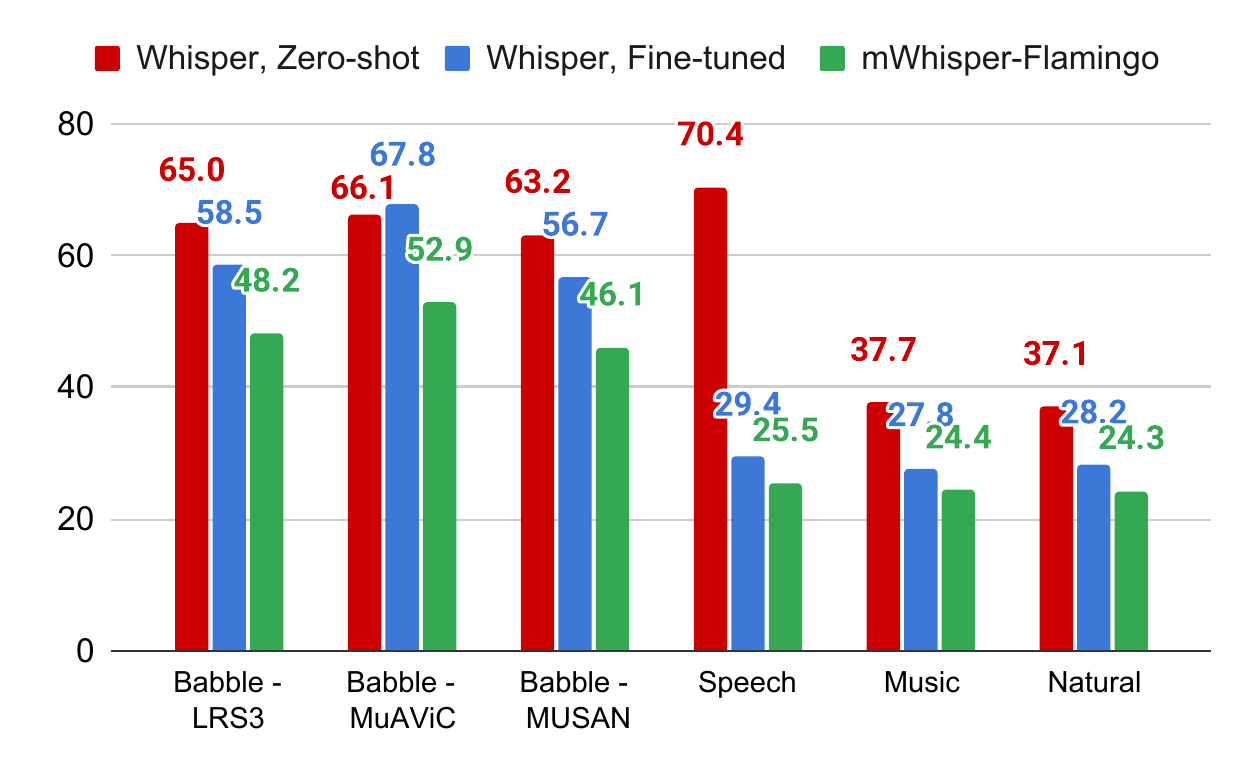}
    \vspace{-2em}
    \caption{Multilingual WER ($\downarrow$ is better) for different noise types averaged over 4 languages (Es, Fr, It, Pt) and 5 SNR levels $\{-10,-5,0,5,10 \}$.}
    \label{fig:noise}
    \vspace{-0.5cm}
\end{figure}

%% file: tables/ablation.tex
\begin{table}[t]
    \centering
    \caption{Ablation Study and Analysis on MuAViC (0db SNR Babble Noise). Performance is WER\%, ($\downarrow$ is better).
    }
    \label{tab:ablation}
    \vspace{-3mm}
\resizebox{\linewidth}{!}{%

\begin{tabular}{lrrrrrr}
\toprule
Setup & \multicolumn{1}{l}{En} & \multicolumn{1}{l}{Es} & \multicolumn{1}{l}{Fr} & \multicolumn{1}{l}{It} & \multicolumn{1}{l}{Pt} & \multicolumn{1}{l}{\begin{tabular}[c]{@{}l@{}}Avg \\ non-En\end{tabular}} \\
\hline
\rowcolor{Gray} \multicolumn{7}{c}{\textit{Whisper-Small Fine-Tuned}} \\
Whisper-Small & 16.2 &41.7 & 35.8 & 50.0 & 50.2 & 44.4 \\
\hline
\rowcolor{Gray} \multicolumn{7}{c}{\textit{Decoder Modality Dropout and Fine-Tunable Visual Encoder}} \\
Mod. Drop. \cmark, FT vis. encoder \cmark & \bf{8.0} & \bf{32.9} & \bf{31.9} & \bf{40.5} & \bf{41.2} & \bf{36.6} \\ 
Mod. Drop. \xmark, FT vis. encoder \cmark & 16.0 & 41.6 & 36.2 & 50.2 & 50.4 & 44.6 \\
Mod. Drop. \xmark, FT vis. encoder \xmark & 11.5 & 39.1 & 35.6 & 47.8 & 47.8 & 42.6 \\
Mod. Drop. \cmark, FT vis. encoder \xmark & 9.7 & 36.7 & 34.1 & 45.1 & 46.5 & 40.6 \\

\hline
\rowcolor{Gray} \multicolumn{7}{c}{\textit{Decoder Modality Dropout Probabilities}} \\
AV: 0.5, A: 0, V: 0.5 & \bf{8.0} & \bf{32.9} & \bf{31.9} & 40.5 & \bf{41.2} & \bf{36.6} \\
AV: 1.0, A: 0, V: 0 & 16.0 & 41.6 & 36.2 & 50.2 & 50.4 & 44.6 \\
AV: 0.5, A: 0.5, V: 0 & 16.0 & 41.6 & 36.1 & 50.3 & 50.5 & 44.6 \\
AV: 0.5, A: 0.25, V: 0.25 & 8.4 & 33.7 & 32.5 & 42.1 & 42.4 & 37.7 \\
AV: 0.25, A: 0.25, V: 0.5 & 8.5 & 34.5 & 32.2 & 41.5 & 42.6 & 37.7 \\
AV: 0.75, A: 0, V: 0.25 & 8.6 & 33.9 & 33.0 & 41.3 & 42.5 & 37.7 \\
AV: 0.25, A: 0, V: 0.75 & 8.3 & 33.5 & 32.7 & \textbf{40.0} & 41.5 & 36.9 \\

\hline
\rowcolor{Gray} \multicolumn{7}{c}{\textit{Multilingual vs English AV-HuBERT}} \\
Multilingual AV-HuBERT & 8.0 & \bf{32.9} & \bf{31.9} & \bf{40.5} & \bf{41.2} & \bf{36.6} \\
English AV-HuBERT & \bf{7.5} & 34.8 & 32.4 & 40.8 & 41.6 & 37.4 \\
\bottomrule
\end{tabular}%
}
\vspace{-0.5cm}
\end{table}

%% file: tables/noise_full_small.tex
\begin{table*}[t]
    \centering
    \setlength{\tabcolsep}{5pt}
    \caption{Multilingual WER ($\downarrow$ is better) on MuAViC with different noise types and SNR levels (small models). The results for Music and Natural noise from MUSAN are averaged.
    }
    \label{tab:noise-full-small}
    \vspace{-3mm}
\resizebox{\linewidth}{!}{%
\begin{tabular}{lrrrrrrrrrrrrrrrrrrrrrrrrr}
\toprule
Method &  \multicolumn{5}{c}{Babble (LRS3), SNR=} & \multicolumn{5}{c}{Babble (MuAViC), SNR=} & \multicolumn{5}{c}{Babble (MUSAN), SNR=} & \multicolumn{5}{c}{Speech (LRS3), SNR=} & \multicolumn{5}{c}{Music+Natural, SNR=} \\
 & -10 & -5 & 0 & 5 & 10 & -10 & -5 & 0 & 5 & 10 & -10 & -5 & 0 & 5 & 10 & -10 & -5 & 0 & 5 & 10 & -10 & -5 & 0 & 5 & 10 \\
 
\hline
\rowcolor{Gray} \multicolumn{26}{l}{\textit{English (En)}} \\
Whisper, Zero-shot & 98.8 & 88.5 & 27.9 & 7.2 & 3.7 & 98.9 & 90.9 & 33.9 & 8.6 & 4.2 & 98.9 & 80.9 & 24.9 & 7.0 & 3.8 & 104 & 86.7 & 36.4 & 8.6 & \textbf{4.0} & 42.7 & 19.3 & 7.5 & 3.9 & \textbf{2.9} \\
Whisper, Fine-tuned & 109 & 79.4 & 16.0 & 5.6 & 4.1 & 120 & 100 & 20.4 & 6.2 & \bf{4.1} & 108 & 63.4 & 13.2 & 4.9 & 4.2 & 74.8 & 51.4 & 26.7 & 12.3 & 6.4 & 36.3 & 13.6 & 5.7 & \bf{4.0} & 3.7 \\
mWhisper-Flamingo & \bf{39} & \bf{27.2} & \bf{8.7} & \bf{4.8} & \bf{4.0} & \bf{39.7} & \bf{32.0} & \bf{10.2} & \bf{5.0} & 4.2 & \bf{38.7} & \bf{24.4} & \bf{7.8} & \bf{4.6} & \bf{4.1} & \bf{41.1} & \bf{26.3} & \bf{13.1} & \bf{6.6} & 4.7 & \bf{15.1} & \bf{7.7} & \bf{4.7} & 4.3 & 4.2 \\

\hline
\rowcolor{Gray} \multicolumn{26}{l}{\textit{Spanish (Es)}} \\
Whisper, Zero-shot & 98.2 & 98.8 & 63.3 & 28.1 & 17.0 & 98.2 & 97.2 & 64.9 & 30.1 & 17.6 & 98.3 & 95.3 & 59.8 & 27.6 & 16.6 & 116 & 111.3 & 76.1 & 30.9 & 16.7 & 62.6 & 40.4 & 24.3 & 16.7 & 13.8 \\
Whisper, Fine-tuned & 108 & 103 & 41.7 & 18.5 & 12.2 & 138 & 117 & 47.5 & 19.8 & 12.9 & 118 & 94.8 & 38.3 & 18.0 & 11.9 & 54.5 & 32.3 & 18.2 & \bf{12.7} & 11.0 & 56.2 & 29.9 & 17.3 & 12.5 & \bf{10.6} \\
mWhisper-Flamingo & \bf{90.1} & \bf{75.7} & \bf{33.6} & \bf{16.9} & \bf{11.7} & \bf{97.4} & \bf{83} & \bf{38.8} & \bf{17.5} & \bf{12.1} & \bf{92.1} & \bf{69.6} & \bf{31.8} & \bf{16.3} & \bf{11.6} & \bf{43.3} & \bf{25.4} & \bf{16.5} & 13.0 & \bf{10.8} & \bf{43.3} & \bf{25.0} & \bf{16.3} & \bf{11.9} & \bf{10.6} \\

\hline
\rowcolor{Gray} \multicolumn{26}{l}{\textit{French (Fr)}} \\
Whisper, Zero-shot & 100 & 96.8 & 57.8 & 31.9 & 22.4 & \textbf{101} & 97.2 & 64.5 & 33.3 & 22.5 & 101 & 95.0 & 54.5 & 30.3 & 22.2 & 111 & 100 & 60.3 & 29.9 & 22.0 & 63.7 & 45.4 & 30.3 & 23.0 & 20.0 \\
Whisper, Fine-tuned & 112 & 90.9 & 35.8 & 20.5 & 16.3 & 144 & 104.0 & 43.3 & 21.7 & 16.0 & 116 & 78.5 & 34.5 & 19.9 & 15.7 & 53.4 & 33.0 & 21.0 & 16.6 & \bf{15} & 53.1 & 31.8 & 21.1 & 15.9 & \bf{14.5} \\
mWhisper-Flamingo & \bf{98.7} & \bf{70.6} & \bf{31.9} & \bf{19.7} & \bf{15.8} & 110 & \bf{83.0} & \bf{39.3} & \bf{20.6} & \bf{15.8} & \bf{98.2} & \bf{63.9} & \bf{31.1} & \bf{19.1} & \bf{15.3} & \bf{44.7} & \bf{28.1} & \bf{19.8} & \bf{16.2} & 15.1 & \bf{44.9} & \bf{27.5} & \bf{19.6} & \bf{15.8} & 14.6 \\

\hline
\rowcolor{Gray} \multicolumn{26}{l}{\textit{Italian (It)}} \\
Whisper, Zero-shot & 99.3 & 101 & 76.2 & 42.7 & 27.7 & 99.8 & 100 & 78.7 & 44.9 & 27.8 & 99.4 & 101 & 71.5 & 39.7 & 26.3 & 112 & 107 & 78.1 & 35.7 & 28.1 & 69.5 & 52.1 & 36.0 & 27.1 & 22.1 \\
Whisper, Fine-tuned & 105 & 102 & 50.6 & 25.6 & 16.6 & 128 & 109 & 53.7 & 25.9 & 16.8 & 110 & 95.0 & 46.9 & 23.9 & 16.4 & 60.7 & 37.2 & 22.9 & 16.5 & \bf{14.4} & 57.5 & 35.1 & 21.9 & 16.2 & 14.1 \\
mWhisper-Flamingo & \bf{86.6} & \bf{75.9} & \bf{41.2} & \bf{22.8} & \bf{16.1} & \bf{94.7} & \bf{81.4} & \bf{43.4} & \bf{23.1} & \bf{16.0} & \bf{87.5} & \bf{71.1} & \bf{37.6} & \bf{22.0} & \bf{15.6} & \bf{48.5} & \bf{31.1} & \bf{21.1} & \bf{16.2} & \bf{14.4} & \bf{46.3} & \bf{30.2} & \bf{19.9} & \bf{15.7} & \bf{13.9} \\

\hline
\rowcolor{Gray} \multicolumn{26}{l}{\textit{Portuguese (Pt)}} \\
Whisper, Zero-shot & 97.8 & 98.9 & 73.4 & 41.3 & 26.9 & \bf{97.5} & 98.6 & 76.2 & 44.0 & 26.5 & 98.0 & 95.3 & 68.8 & 38.8 & 25.5 & 110 & 108 & 86.2 & 43.9 & 25 & 67.4 & 51.8 & 34.8 & 25.9 & 21.2 \\
Whisper, Fine-tuned & 109 & 107 & 50.2 & 26.9 & 18.0 & 140 & 116 & 56.3 & 28.2 & 18.0 & 114 & 96.3 & 45.5 & 24.2 & 17.5 & 67.3 & 41.5 & 25.1 & 18.7 & 15.7 & 60.7 & 36.5 & 23.0 & 17.4 & 15.0 \\
mWhisper-Flamingo & \bf{91.3} & \bf{80.9} & \bf{42.8} & \bf{23.9} & \bf{17.2} & 104 & \bf{89.8} & \bf{45.7} & \bf{25.0} & \bf{17.5} & \bf{90.1} & \bf{71.9} & \bf{37.7} & \bf{22.8} & \bf{16.7} & \bf{53.0} & \bf{35.2} & \bf{23.4} & \bf{18.3} & \bf{15.5} & \bf{48.1} & \bf{31.1} & \bf{21.0} & \bf{16.7} & \bf{14.6} \\

\bottomrule
\end{tabular}%
}
\end{table*}

%% file: tables/noise_full_medium.tex
\begin{table*}[t]
    \centering
    \setlength{\tabcolsep}{5pt}
    \caption{Multilingual WER ($\downarrow$ is better) on MuAViC with different noise types and SNR levels (medium models). The results for Music and Natural noise from MUSAN are averaged.
    }
    \label{tab:noise-full-medium}
    \vspace{-3mm}
\resizebox{\linewidth}{!}{%
\begin{tabular}{lrrrrrrrrrrrrrrrrrrrrrrrrr}
\toprule
Method &  \multicolumn{5}{c}{Babble (LRS3), SNR=} & \multicolumn{5}{c}{Babble (MuAViC), SNR=} & \multicolumn{5}{c}{Babble (MUSAN), SNR=} & \multicolumn{5}{c}{Speech (LRS3), SNR=} & \multicolumn{5}{c}{Music+Natural, SNR=} \\
 & -10 & -5 & 0 & 5 & 10 & -10 & -5 & 0 & 5 & 10 & -10 & -5 & 0 & 5 & 10 & -10 & -5 & 0 & 5 & 10 & -10 & -5 & 0 & 5 & 10 \\
 
\hline
\rowcolor{Gray} \multicolumn{26}{l}{\textit{English (En)}} \\
Whisper, Zero-shot & 98.9 & 84.0 & 22.5 & 5.6 & \textbf{2.6} & 99.7 & 86.1 & 26.6 & 6.6 & 3.0 & 99.2 & 74.0 & 20.1 & 5.5 & \textbf{2.8} & 99.2 & 74.7 & 29.2 & \textbf{5.9} & \textbf{2.8} & 37.8 & 16.0 & 5.8 & \bf{3.1} & \bf{2.4} \\
Whisper, Fine-tuned & 110 & 72.3 & 12.3 & 4.5 & 3.6 & 116 & 91.6 & 16.6 & 5.0 & 3.8 & 108 & 55.9 & 11.4 & 4.2 & 3.5 & 59.7 & 40.3 & 23.4 & 12.7 & 7.5 & 32.6 & 11.4 & 5.0 & 4.1 & 3.8 \\
mWhisper-Flamingo & \bf{40.0} & \bf{25.7} & \bf{7.5} & \bf{3.8} & 3.4 & \bf{40.9} & \bf{32.2} & \bf{8.7} & \bf{4.0} & \bf{3.6} & \bf{39.7} & \bf{22.1} & \bf{6.1} & \bf{3.7} & 3.6 & \bf{34.4} & \bf{23.1} & \bf{13.5} & 7.2 & 4.7 & \bf{13.5} & \bf{6.4} & \bf{4.1} & 3.5 & 3.5 \\

\hline
\rowcolor{Gray} \multicolumn{26}{l}{\textit{Spanish (Es)}} \\
Whisper, Zero-shot & 99.7 & 92.6 & 49.1 & 20.7 & 12.6 & 99.8 & 95.0 & 52.2 & 21.2 & 13.1 & 100.8 & 88.2 & 46.3 & 20.1 & 12.5 & 112.3 & 92.7 & 49.8 & 19.2 & 13.5 & 55.6 & 33.4 & 18.8 & 12.8 & 10.8 \\
Whisper, Fine-tuned & 111 & 97.0 & 36.1 & 14.9 & 11.3 & 129 & 108 & 38.5 & 15.4 & 11.4 & 114.9 & 83.2 & 31.1 & 14.1 & \bf{9.9} & 37.3 & 20.2 & 13.1 & 10.4 & \bf{8.9} & 47.6 & 25.6 & 13.6 & 10.5 & \bf{8.8} \\
mWhisper-Flamingo & \bf{91.2} & \bf{70.3} & \bf{28.0} & \bf{13.9} & \bf{9.9} & \bf{97.6} & \bf{77.8} & \bf{32.2} & \bf{14.4} & \bf{10.1} & \bf{91.0} & \bf{63.1} & \bf{26.2} & \bf{13.1} & 10.4 & \bf{31.0} & \bf{18.0} & \bf{12.1} & \bf{10.0} & 9.0 & \bf{38.6} & \bf{20.3} & \bf{12.5} & \bf{9.8} & \bf{8.8} \\

\hline
\rowcolor{Gray} \multicolumn{26}{l}{\textit{French (Fr)}} \\
Whisper, Zero-shot & 100 & 90.2 & 47.7 & 25.1 & 19.3 & \bf{99.9} & 93.9 & 52.2 & 27.1 & 18.7 & 100 & 85.5 & 44.0 & 24.5 & 19.3 & 103.3 & 73.6 & 37.2 & 23.0 & 19.4 & 56.4 & 36.2 & 24.6 & 18.9 & 16.6 \\
Whisper, Fine-tuned & 116 & 78.2 & 30.4 & 16.2 & \bf{12.4} & 139 & 97.6 & 35.6 & 17.5 & \bf{12.6} & 118 & 72.7 & 28.4 & 15.6 & \bf{12.4} & 36.9 & 22.9 & \bf{15.8} & \bf{12.9} & 12.0 & 47.4 & 25.4 & 16.0 & \bf{13.0} & \bf{11.8} \\
mWhisper-Flamingo & \bf{98.1} & \bf{63.3} & \bf{27.5} & \bf{16.0} & 12.7 & 107 & \bf{75.1} & \bf{30.7} & \bf{17.1} & 12.7 & \bf{97.7} & \bf{58.3} & \bf{26.2} & \bf{15.5} & 12.7 & \bf{32.4} & \bf{21.0} & 15.9 & 13.2 & \bf{11.9} & \bf{39.3} & \bf{23.1} & \bf{15.8} & \bf{13.0} & 12.1 \\

\hline
\rowcolor{Gray} \multicolumn{26}{l}{\textit{Italian (It)}} \\
Whisper, Zero-shot & 99.9 & 96.8 & 60.3 & 28.9 & 16.8 & 99.9 & 99.6 & 62.6 & 29.8 & 17.2 & 100 & 93.3 & 55.7 & 27.2 & 16.1 & 110.1 & 96.2 & 53.6 & 24.1 & 15.0 & 59.6 & 39.2 & 23.2 & 16.1 & 13.3 \\
Whisper, Fine-tuned & 112 & 102 & 43.5 & 19.8 & 13.0 & 121 & 102 & 47.2 & 20.0 & 12.9 & 117 & 87.9 & 39.3 & 18.7 & 12.6 & 44.5 & 25.3 & 16.6 & 13.1 & 11.5 & 52.1 & 29.6 & 17.5 & 12.8 & \bf{11.2} \\
mWhisper-Flamingo & \bf{92.7} & \bf{73.3} & \bf{35.2} & \bf{18.0} & \bf{12.6} & \bf{94.6} & \bf{77.4} & \bf{37.0} & \bf{18.3} & \bf{12.6} & \bf{89.8} & \bf{66.5} & \bf{31.6} & \bf{17.2} & \bf{12.1} & \bf{35.4} & \bf{22.2} & \bf{15.5} & \bf{12.5} & \bf{11.2} & \bf{42.6} & \bf{24.8} & \bf{15.9} & \bf{12.6} & 11.4 \\

\hline
\rowcolor{Gray} \multicolumn{26}{l}{\textit{Portuguese (Pt)}} \\
Whisper, Zero-shot & 97.0 & 93.2 & 60.9 & 31.2 & 18.9 & \textbf{97.1} & 94.8 & 63.1 & 32.1 & 19.5 & 97.0 & 90.7 & 55.7 & 28.4 & 18.5 & 109 & 98.9 & 69.2 & 31.9 & 18.9 & 60.1 & 41.1 & 26.1 & 18.3 & 15.2 \\
Whisper, Fine-tuned & 109 & 95.4 & 42.3 & 21.7 & 13.9 & 130 & 104 & 46.3 & 21.5 & 14.2 & 114.4 & 85.1 & 37.7 & 19.4 & 13.7 & 45.4 & 27.9 & 18.2 & 14.3 & 12.4 & 52.3 & 30.3 & 18.7 & 14.0 & 12.2 \\
mWhisper-Flamingo & \bf{94.5} & \bf{74.3} & \bf{36} & \bf{20.0} & \bf{13.7} & 100 & \bf{79.7} & \bf{38.6} & \bf{20.1} & \bf{14.1} & \bf{93.6} & \bf{65.4} & \bf{32.8} & \bf{18.2} & \bf{13.5} & \bf{38.1} & \bf{25.0} & \bf{17.2} & \bf{13.8} & \bf{12.3} & \bf{43.4} & \bf{26.3} & \bf{17.4} & \bf{13.5} & \bf{11.9} \\

\bottomrule
\end{tabular}%
}
\end{table*}